\documentclass[]{article}
\usepackage{arxiv}  
\usepackage{natbib}
\usepackage{times}  
\usepackage{helvet}  
\usepackage{courier}  
\usepackage[hyphens]{url}  
\usepackage{caption} 
\usepackage{algorithm}
\usepackage{algorithmic}
\usepackage{newfloat}
\usepackage{listings}
\usepackage[utf8]{inputenc} 
\usepackage{url}            
\usepackage{booktabs}       
\usepackage{amsfonts}       
\usepackage{nicefrac}       
\usepackage{microtype}      
\usepackage{lipsum}		
\usepackage{graphicx}
\usepackage{doi}
\usepackage[toc,title,page]{appendix}
\usepackage{subcaption}
\usepackage{numprint}
\usepackage{tikz}
\usepackage{xcolor}
\usetikzlibrary{shapes.geometric, arrows, fit}

\newcommand{\diffp}{\textsc{DP}}
\newcommand{\fsha}{\textsc{FSHA}}
\newcommand{\mnist}{\textsc{MNIST}}
\newcommand{\fashionmnist}{\textsc{Fashion-MNIST}}
\newcommand{\celeb}{\textsc{CelebA}}
\newcommand{\finv}{\ensuremath{\tilde{f}^{-1}}}
\newcommand{\ftilde}{\ensuremath{\tilde{f}}}
\newcommand{\f}{\ensuremath{f}}

\newcommand{\pca}{\textsc{PCA}}

\floatstyle{ruled}
\title{Feature Space Hijacking Attacks against Differentially Private Split Learning}

\author{
  Grzegorz Gawron \\
  LiveRamp, Privacy Tech R\&D \\
  \texttt{greg.gawron@liveramp.com}
  \And
  Philip Stubbings \\
  LiveRamp, Privacy Tech R\&D \\
  \texttt{phil.stubbings@liveramp.com}
}

\hypersetup{
    pdftitle={Feature Space Hijacking Attacks against Differentially Private Split Learning},
    pdfauthor={Grzegorz Gawron, Phil Stubbings},
    pdfkeywords={Neural Networks, Split Learning, differential privacy, privacy attacks},
}

\begin{document}

\maketitle

\begin{abstract}

Split learning and differential privacy are technologies with growing potential to help with privacy-compliant advanced analytics on distributed datasets.
Attacks against split learning are an important evaluation tool and have been receiving increased research attention recently.
This work's contribution is applying a recent feature space hijacking attack (\fsha{}) to the learning process of a split neural network enhanced with differential privacy (\diffp{}), using a client-side off-the-shelf DP optimizer. 
The \fsha{} attack obtains client's private data reconstruction with low error rates at arbitrarily set \diffp{} $\epsilon$ levels. 
We also experiment with dimensionality reduction as a potential attack risk mitigation and show that it might help to some extent. We discuss the reasons why differential privacy is not an effective protection in this setting and mention potential other risk mitigation methods.

\end{abstract}

\section{Introduction}
\label{sec:introduction}

Privacy compliant data processing has been receiving an increasing amount of attention recently. 
Take a consortium of hospitals as a motivational example. Each owns a data set describing treatments given to patients, together with the treatments' outcomes.
There is a potential great value in learning from all of the data sets taken together. However, the hospitals are bound by privacy regulations that prevent them from sharing the data in raw form. For a thorough discussion of distributed learning from patient data see \citet{vepakomma2018split}.

Split learning (see \citet{gupta2018distributed}), a type of federated learning, has been proposed to address scenarios like this. It is a neural network training mechanism, which \textbf{splits} the network into modules corresponding to the private data silos, so that each module is trained within the silo and no raw data is exposed outside. 
Another reason for growing popularity of split learning is it's performance. Since only the processed or \textbf{smashed} data of lower dimensions is being exchanged among silos and their coordinator, the process is more efficient than some of the more established mechanisms of distributed learning, such as federated averaging.
Yet, relying on smashing the data as the only mechanism of protecting privacy has been shown by \citet{Erdogan2021upsplit} to be misguided and that even an honest-but-curious actor having access to smashed data can reconstruct the private inputs.
\citet{vepakomma2018peek} have surveyed various methods to deal with split learning privacy leaks having, what they call a \textbf{no peek} learning, as their target. 

\begin{figure}[ht]
  \centering
  \begin{center}
    \includegraphics[width=8cm]{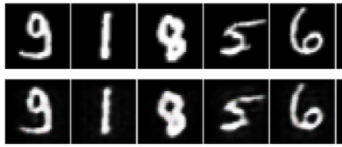}
  \end{center}
    \caption{Plots comparing the private input (top row) and its corresponding reconstruction (bottom row) in a successful \fsha{} attack on \mnist{}. }
\label{fig:orig-10k}
\end{figure}

\begin{figure}[ht]
    \centering
    \includegraphics[width=8cm]{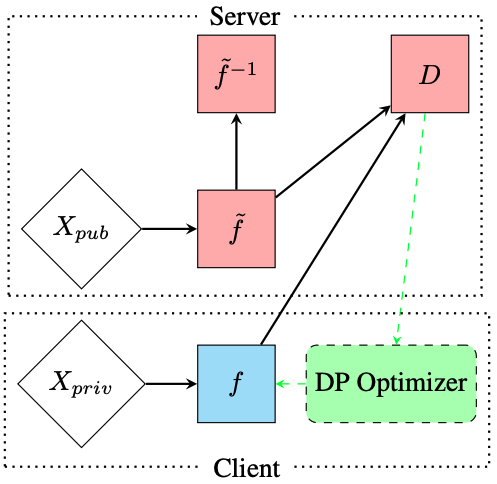}
    \caption{FSHA split network architecture with the \diffp{} logic implemented by a DPOptimizer from \textsc{Tensorflow-Privacy} package}
\label{fig:fsha-architecture}
\end{figure}

In this paper we focus on another technique which can be used to enhance split learning's privacy compliance: differential privacy (\diffp{}). Its promise is to provide a rigorous mathematical privacy guarantee (see \citet{Dwork2014}). \citet{Abadi2016} showed that training with \diffp{} is feasible, while having control over model performance degradation and privacy cost. There has also been work by \citet{papernot2020making} on fine tuning deep learning with \diffp{} to minimise the performance cost with some good results.
\citet{geyer2017differentially} is another example of the work on optimizing the \diffp{} deep learning process.

Theoretical privacy guarantees, when applied in practice, need to be backed by automatic evaluation, which helps to guard against plain bugs or other implementation issues.
There are multiple, already well established libraries available, which provide the necessary evaluation components, e.g. \textsc{Adversarial-Robustness-Toolbox}\footnote{\url{https://github.com/Trusted-AI/adversarial-robustness-toolbox}} or \textsc{Tensorflow-Privacy}\footnote{\url{https://github.com/tensorflow/privacy}}. For a comprehensive overview of attack and defences literature and evaluation best practices we refer the reader to \citet{carlini2019evaluating}.

Attack based evaluation is particularly relevant for newly emerging methods like split learning. There have already been first papers published on this subject.
One recent attack method is feature space hijacking attack \fsha{} by \citet{pasquini2021unleashing}, where a malicious attacker taking over the server module is able to  reconstruct all of the client private input data with arbitrarily low reconstruction error on \textsc{MNIST} and other visual datasets.
See Figure~\ref{fig:orig-10k} for an illustration of a successful \fsha{} attack on a \mnist{} dataset, based on an experiment reproduction from the original paper - without \diffp{} applied.

As a testament to the rapid pace of research progress in the field, methods guarding against such attacks have already been presented, see \citet{Erdogan2021splitguard}.

\textbf{Our contributions.}  
This work contributes by applying differential privacy to the client module of a split neural network, running a suite of \fsha{} attacks against it and drawing conclusions for differentially private split learning system implementations. 
To our knowledge this is the first FSHA style attack attempt at such a setup. 
\citet{Hitaj2017} have shown the limits of applying \diffp{} to general collaborative deep learning and we validate the vulnerability of split learning in that respect. 

We reuse the software prototype from \citet{pasquini2021unleashing} found at \url{https://github.com/pasquini-dario/SplitNN_FSHA} and aim to introduce minimal code changes to reach our goal. We also follow the authors' choice of using \textsc{Tensorflow} and use its sub-package \textsc{Tensorflow-Privacy} for the application of \diffp{} clipping and noise. See \citet{mcmahan2019general} for the description of this package.

\section{Architecture}

In what follows we briefly describe the original \fsha{} attack split learning architecture and refer the reader to the original paper of \citet{pasquini2021unleashing} for a complete discussion of the components (we use the same notation as in the original paper).
We restrict our investigation to a situation where the attacker takes over only the server. The clients are not taken over and believe they are part of a legitimate cooperation. However, as highlighted in the original paper, the same attack would work successfully by taking over one of the clients only. 
The example neural network in Figure~\ref{fig:fsha-architecture} is split into two main components: client and server (in general there could be multiple clients). 
The client owns their private data stored in a silo $X_{priv}$ and is responsible for training the $f$ model component of the network.
The server runs the attacker's code, which instead of running its part of the split learning protocol, trains two components: discriminator $D$ and an auto-encoder $\ftilde$ and $\finv$. The auto-encoder, when given inputs from $X_{pub}$, encodes them using $\ftilde$ to an internal representation (feature space) and decodes using $\finv$ with the loss set, so as to learn to reconstruct the original input data. Finally, discriminator $D$ is responsible for classifying the input as coming from either $f$ model or the $\ftilde$ model.
The ultimate attacker's target is for the discriminator $D$ to force the private model $f$'s output to come from the same feature space as the outputs of $\ftilde$ and the $\finv$ to be able to decode it with minimal error, just as it learned to decode the $\ftilde$.

The standard way of applying differential privacy during neural network training is on the gradients during back propagation.
In our scenario the most natural place to apply the \diffp{} clipping and noise to the $\f$ gradients is just before applying the gradient updates sent back by server's discriminator $D$ as illustrated by the green fragments in Figure~\ref{fig:fsha-architecture}.
This allows the client data owner to control the noise application process and configure it according to their privacy preferences.

\section{Experiments}
    
\subsection{Experimental Setup}

We enhanced the FSHA code with DP by introducing only minimal changes, with the aim of being able to easily compare the results. We used \textbf{TensorFlow-privacy} python package to add a differentially private optimizer \textbf{DPAdamGaussianOptimizer}. We parametrised DP with $\delta=0$ and various levels of $\epsilon$ from $1$ to $100$.

We deployed several GCP Notebooks instances with the following hardware: (a)  a2-highgpu-1g (Accelerator Optimized: 1 NVIDIA Tesla A100 GPU, 12 vCPUs, 85GB RAM); and (b) n1-standard-4 (4 vCPUs, 15 GB RAM, 1 NVIDIA Tesla T4).
GPU load was $30$ percent on (a) and $80$ percent on (b) with similar run times, so we settled on the latter to optimize the experiment cost.

\begin{figure}[ht]
    \begin{subfigure}{0.5\textwidth}
        \centering
        \includegraphics[width=8cm]{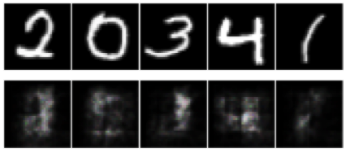}
        \caption{\numprint{10000} iterations, DP with $\epsilon=10$.}
        \label{fig:eps10-10k}
    \end{subfigure}\hfill
    \begin{subfigure}{0.5\textwidth}
        \centering
        \includegraphics[width=8cm]{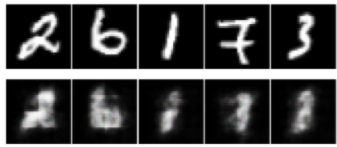}
        \caption{\numprint{10000} iterations, DP with $\epsilon=100$.}
        \label{fig:eps100-10k}
    \end{subfigure}\hfill
    \caption{Private input (top row) and its corresponding reconstruction (bottom row) for various epsilon settings and \numprint{10000} iterations.} 
\end{figure}

\subsection{\mnist{} Results}

We start with the \textsc{MNIST} dataset because it is widely used for benchmarking and its simplicity helps to clearly show the impact of the application of \diffp. Note, that we focus on showing the reconstruction visualisations, as they nicely communicate the results. The diagrams of the actual reconstruction errors can be found in the Appendix.

\begin{figure}[ht]
    \begin{subfigure}{0.5\textwidth}
        \centering
        \includegraphics[width=8cm]{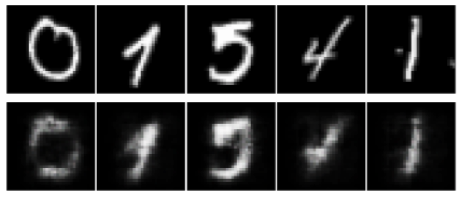}
        \caption{\numprint{100000} iterations, DP with $\epsilon=0.5$.}
        \label{fig:eps0.5-100k}
    \end{subfigure}\hfill
    \begin{subfigure}{0.5\textwidth}
        \centering
        \includegraphics[width=8cm]{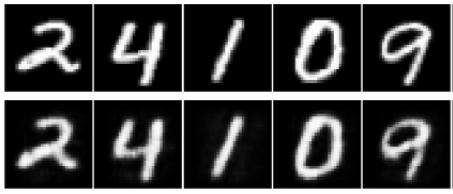}
        \caption{\numprint{100000} iterations, DP with $\epsilon=10$.}
        \label{fig:eps10-100k}        
    \end{subfigure}\hfill    
    \caption{Plots comparing the private input (top row) and its corresponding reconstruction (bottom row) for various epsilon settings and \numprint{100000} iterations.} 
\end{figure}

We reproduced the results from the original paper, without any differential privacy applied. Figure~\ref{fig:orig-10k} shows a successful \fsha{} reconstruction attack after \numprint{10000} iterations of $64$ element mini-batches. This corresponds to $10$ epochs of training, since there are $N=\numprint{60000}$ samples in the training set.
The next step is to run the same number of iterations using DP with $\epsilon=10$ and $\delta=1/N$. Figure~\ref{fig:eps10-10k} shows that the reconstruction was far less successful, however the reconstruction error was still declining after the training stopped. This suggested that more iterations might make the attack successful. As expected the attack against a split learning with $\epsilon=100$ is able to reconstruct the private data with much less error (Figure~\ref{fig:eps100-10k}).
We continued with the training process running for \numprint{100000} iterations ($100$ epochs) and at $\epsilon=10$ it was enough to produce very low-error reconstructions, as illustrated in Figure~\ref{fig:eps10-100k}.
Finally, we set $\epsilon$ to a much lower value $0.5$ and obtained the reconstruction results shown in Figure~\ref{fig:eps0.5-100k}. Although the error is much higher than with the previous experiment (where $\epsilon=10$) the reconstruction error was still steadily declining, so by increasing the number of iterations we could expect the error to go down as per previous experiments.

See the Appendix for a more complete record of reconstruction plots.

\subsection{\fashionmnist{} Results}

Beyond the \fashionmnist{} attacks reproduced from \citet{pasquini2021unleashing} we have experimented with removing some of the image classes from the public data set and applying differential privacy in both settings.

\begin{figure}[ht]
    \centering
    \begin{subfigure}{0.45\textwidth}
        \centering
        \includegraphics[width=8cm]{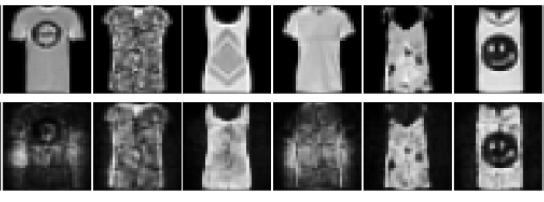}
        \caption{no DP, iterations $=\numprint{100000}$}
        \label{fig:fmnist-epsNone-it100k}
    \end{subfigure}\hfill
    \begin{subfigure}{0.45\textwidth}
        \centering
        \includegraphics[width=8cm]{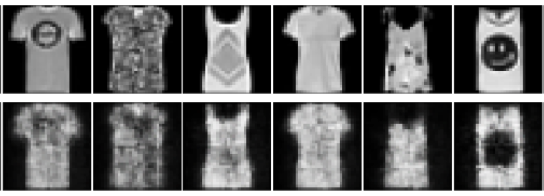}
        \caption{$\epsilon=10$, iterations $=\numprint{100000}$}
        \label{fig:fmnist-eps10-it100k}
    \end{subfigure}
    
    \begin{subfigure}{0.45\textwidth}
        \centering
        \includegraphics[width=8cm]{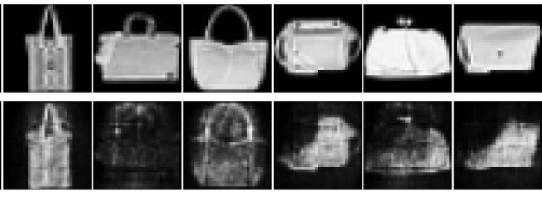}
        \caption{no DP, iterations $=\numprint{100000}$}
        \label{fig:fmnist-remcat=8,eps=None,its=100k}
    \end{subfigure}\hfill
    \begin{subfigure}{0.45\textwidth}
        \centering
        \includegraphics[width=8cm]{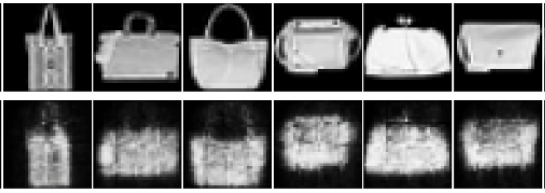}
        \caption{$\epsilon=10$, iterations $=\numprint{100000}$}
        \label{fig:fmnist-remcat=8,eps=10,its=100k}
    \end{subfigure}
    \hfill
    \caption{FSHA against \fashionmnist{} with category $0$ (tops) missing from the public data set in Figures~\ref{fig:fmnist-epsNone-it100k} and \ref{fig:fmnist-eps10-it100k} and category $8$ (bags) missing in Figures~\ref{fig:fmnist-remcat=8,eps=None,its=100k} and \ref{fig:fmnist-remcat=8,eps=10,its=100k}.}
\end{figure}

We first removed the category $0$ representing t-shirts/tops and ran the attacks with and without the \diffp{}. We present the relevant results in Figure~\ref{fig:fmnist-epsNone-it100k} and Figure~\ref{fig:fmnist-eps10-it100k}. In both cases the shapes were mostly visible, but the DP version had less logo details visible (see the two pictures to the right). 

There are, however, other categories in the public data set (see Figure~\ref{fig:fmnist-cats}) that are similar to the t-shirts, so the process had more chance to learn the relevant features to also reconstruct t-shirts.
In an attempt to minimise this effect, we remove the category which is least similar to others: bags (category = $8$). The result is illustrated in Figures \ref{fig:fmnist-remcat=8,eps=None,its=100k} and \ref{fig:fmnist-remcat=8,eps=10,its=100k}. Interestingly, the version without \diffp{} reconstructs some of the bags as shoes, perhaps because they activate similar latent features learned by the attack model. However we can recognise certain features coming from the private data e.g., the circle on the right t-shirt. 
With \diffp{} added we see no 'bag as shoe' effect. The images are more blurred, but the general shapes are reconstructed. This might be due to regularisation effect of DP.
It is important to note that it should be possible to improve this results by running more training iterations (we stopped at \numprint{100000}).

To summarise, \fashionmnist{} data set with some of the classes removed from the public data set achieves lower reconstruction precision, especially with \diffp{} applied. This helps demonstrate the point that the reconstructions are really generated out of the feature space driven by the access to similar public data. The closer the public and private data distributions are, the more successful the attacks.

\subsection{Mitigating the attack with dimensionality reduction}

We explored the application of \diffp{} during the model training process as a means to
mitigate the attack and demonstrated that DP can at most delay \fsha{} convergence.

In an un-trusted environment in which it can be assumed a client or server will
attempt to exploit the learning protocol by means of the \fsha{} method, additional
steps should be taken to protect the underlying data. One possibility is to
apply dimensionality reduction techniques directly to the data prior to training
to a level which still yields acceptable model accuracy.

\begin{figure}[ht]
    \begin{subfigure}{0.5\textwidth}
    \centering
    \includegraphics[width=.95\textwidth]{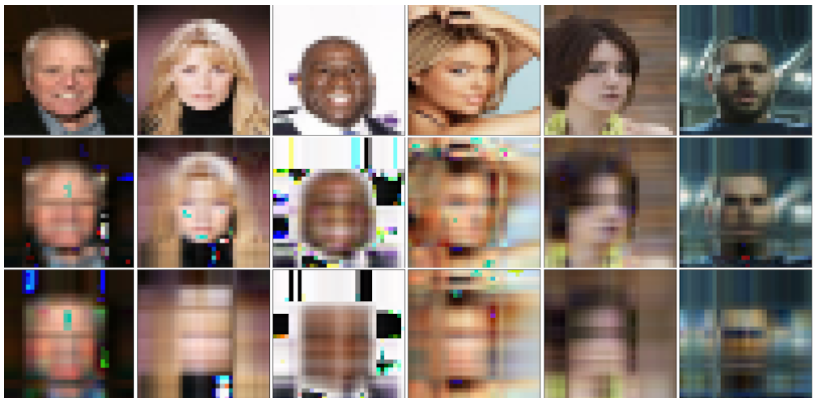}
    \caption{Image compression}
    \label{fig:raw_3}
    \end{subfigure}\hfill
    \centering
    \begin{subfigure}{0.5\textwidth}
    \includegraphics[width=0.95\textwidth]{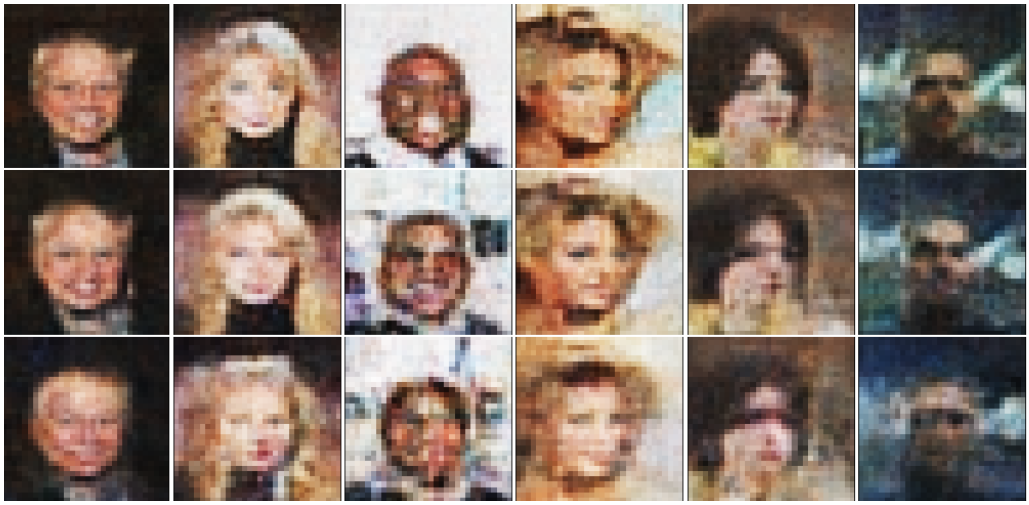}
    \caption{\fsha{} recovered images}
    \label{fig:recovered_3}
    \end{subfigure}%
\caption{PCA image compression and recovery. $1^{st}$ row - no compression, $2^{nd}$ row - principal components = 4, $3^{rd}$ row - principal components = 2.}
\end{figure}

To test this hypothesis, we re-run the \fsha{} attack up to 100k iterations on image
data which has been compressed using various levels of \pca{} principal components as
shown in Figure~\ref{fig:raw_3} (also in Appendix Figure~\ref{fig:raw_6}). We make use of the \celeb{} dataset and define the
classification task to predict whether the celebrity is likely male or female.

The resulting recovered images shown in Figure~\ref{fig:recovered_3} (also in Appendix Figure~\ref{fig:recovered_6}) demonstrate
that applying dimensionality reduction to the data can mitigate the attack to a
degree in which the celebrities are no longer identifiable. 
One should note though, that some reconstructed image features might still be obvious and unique to certain individuals.

\section{Conclusions}

The \fsha{} mechanism used is very robust. It only requires that the client part of the model can be successfully trained; and for any useful usage that must be the case. Once the attacker obtains a sufficiently good client model they can reconstruct the private data which are passed from the client to server during the training process. As \citet{pasquini2021unleashing} notice, they could also infer certain features out of the private data, if that's what they choose to do.

It might seem surprising that the \diffp{} doesn't give enough protection against \fsha{}. However, the \diffp{} noise applied during the training is designed to prevent the model from memorizing data so that the private data cannot be retrieved from the trained model itself. It must still allow for training useful models. After a successful training, a good \diffp{} model must be able to achieve high inference accuracy. The inferred values might well be sensitive and constitute a privacy leak. Hence, having access to the model being trained allows an attacker to either reconstruct or infer properties of the private data. We refer the reader to \citet{Hitaj2017} for a more thorough discussion of this topic.

Employing some dimensionality reduction techniques might make the potential reconstructions useless, but we must remember that it might not always produce a model with good enough utility and even more importantly the attacker might still infer sensitive properties from the private data.

Detection methods such as ones proposed by \citet{Erdogan2021splitguard} might also be used, but they are not bullet proof and require constant attention to new detection method workarounds used by potential attackers.

As for now, to ensure that the split learning process is safe we should mitigate the risk of someone taking over the server (or indeed any other split network component). 
The risk mitigation might rely on enforcing a secure compute environment with e.g. only a certified version of the training code running on all the modules of the split network. 
It also follows that the application of the \diffp{} trained split neural network model should happen in an environment of trust. That is, the owners of private data should trust that inferred outputs are appropriately taken care of. It is safe, for instance, if the model is applied directly by the data owner on their own data and they own the inferred results. 

\section*{Acknowledgements}
This work has benefited from valuable input during the 2021 OpenDP Fellows Program (see \url{https://opendp.org/fellows-program}).

\bibliography{references}

\appendix
\section{Appendix}
Below we include the complete figures from our experiments, where they didn't fit in the main body of the paper.

\begin{figure*}[ht]
    \begin{subfigure}{0.45\textwidth}
        \raggedleft
        \includegraphics[width=.8\textwidth]{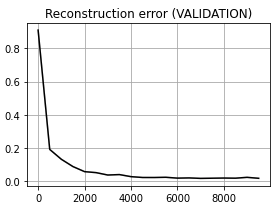}
        \caption{\numprint{10000} iterations, no DP applied (original reconstruction).}
        
    \end{subfigure}\hfill        
    \begin{subfigure}{0.45\textwidth}
        \raggedright
        \includegraphics[width=.8\textwidth]{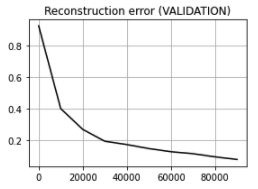}
        \caption{\numprint{100000} iterations, DP with $\epsilon=0.5$.}
        
    \end{subfigure}\hfill
    \begin{subfigure}{.45\textwidth}
        \raggedleft
        \includegraphics[width=.8\textwidth]{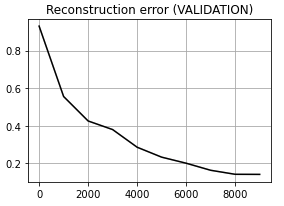}
        \caption{\numprint{10000} iterations, DP with $\epsilon=100$.}
        
    \end{subfigure}
    \begin{subfigure}{0.45\textwidth}
        \raggedright
        \includegraphics[width=.8\textwidth]{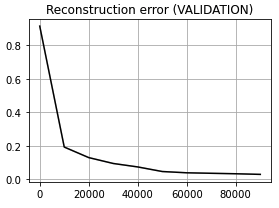}
        \caption{\numprint{100000} iterations, DP with $\epsilon=10$.}
        
    \end{subfigure}\hfill    
    \begin{subfigure}{.45\textwidth}
        \raggedleft
        \includegraphics[width=.8\textwidth]{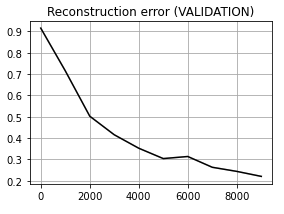}
        \caption{\numprint{10000} iterations, DP with $\epsilon=10$.}
        
    \end{subfigure}\hfill
    \caption{Reconstruction errors (vertical axis) against iterations (horizontal axis) for various iteration levels and DP application parameters. } 
\end{figure*}

\begin{figure*}[ht]
\centering
\includegraphics[width=0.95\textwidth]{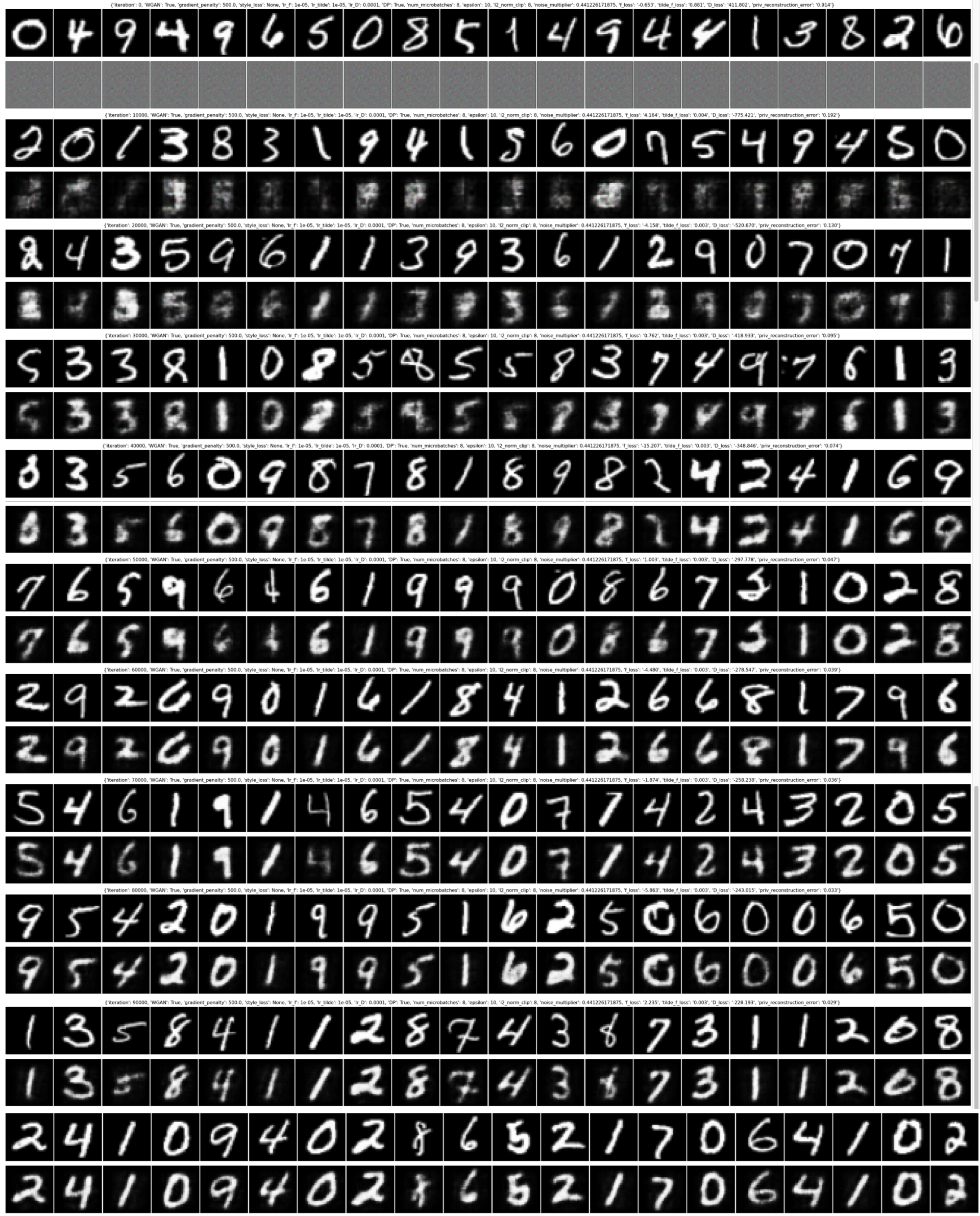}
\caption{Reconstruction attempts for $\epsilon=10$ and \numprint{100000} iterations. Odd rows show the private data fed to the private local $f$ model and the even rows below show the corresponding images reconstructed using attacker's $f^{-1}$ model. The lower the line, the more training iterations there were.}
\label{fig:mnist-eps=10-100k-iterations-full}
\end{figure*}

\begin{figure*}[ht]
\centering
\includegraphics[width=0.95\textwidth]{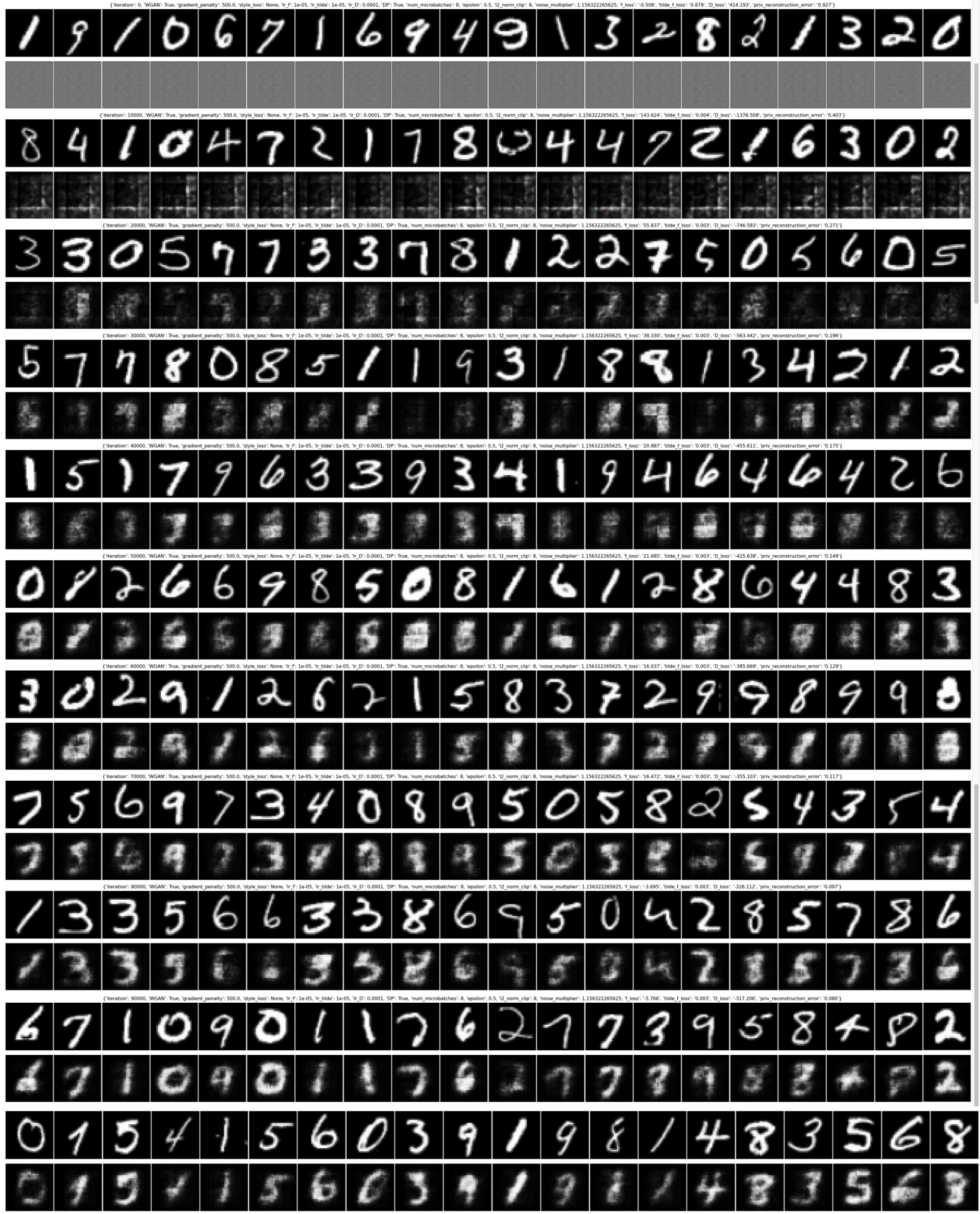}
\caption{Reconstruction attempts for $\epsilon=0.5$ and \numprint{100000} iterations. Odd rows (with little titles) show the private data fed to the private local $f$ model and the even rows below show the corresponding images reconstructed using attacker's $f^{-1}$ model. The lower the line, the more training iterations there were.}
\label{fig:mnist-eps=0.5-100k-iterations-full}
\end{figure*}

\begin{figure*}[ht]
\centering
\includegraphics[width=8cm]{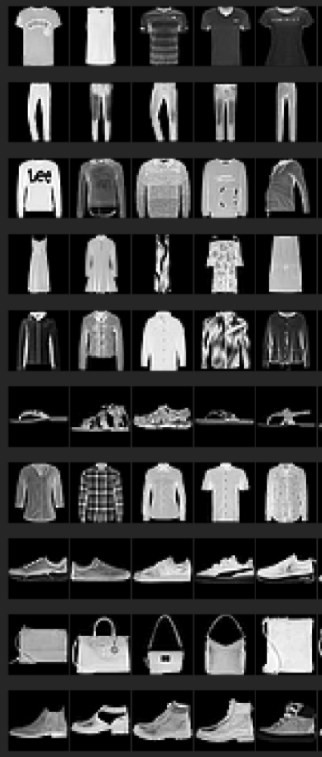}
\caption{A few examples per each of the ten \fashionmnist{} categories}
\label{fig:fmnist-cats}
\end{figure*}

\begin{figure*}[ht]
\centering
    \begin{subfigure}{.7\textwidth}
    \centering
    \includegraphics[width=.8\textwidth]{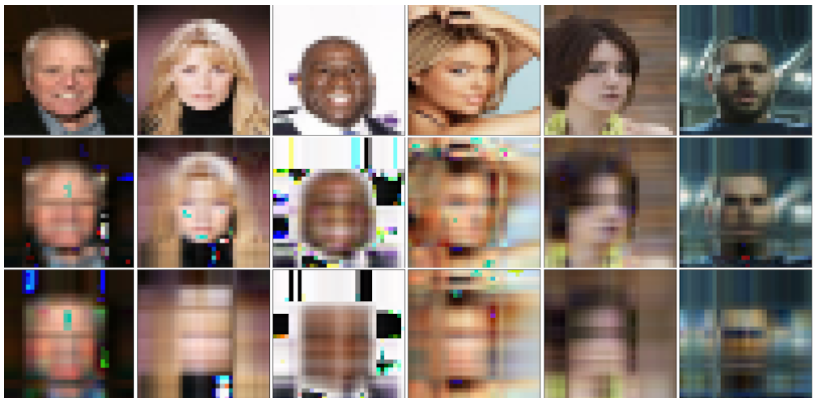}
    \caption{Image compression}
    \label{fig:raw_6}
    \end{subfigure}
    \begin{subfigure}{.7\textwidth}
    \centering
    \includegraphics[width=0.8\textwidth]{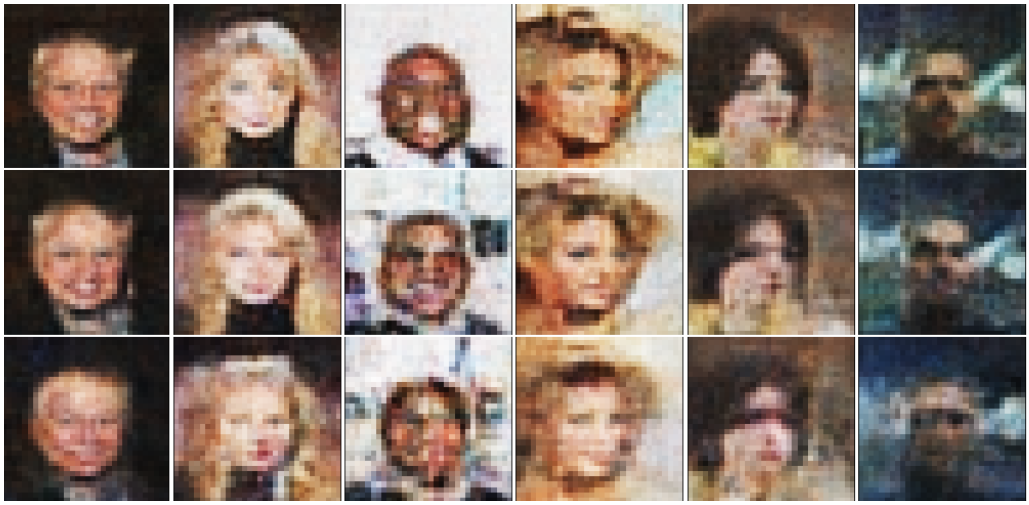}
    \caption{\fsha{} recovered images}
    \label{fig:recovered_6}
    \end{subfigure}
\caption{PCA image compression and recovery. $1^{st}$ row - no compression, $2^{nd}$ row - principal components = 4, $3^{rd}$ row - principal components = 2.}
\end{figure*}

\end{document}